# Study of the *d(p,γ)*³He Reaction at Ultralow Energies using a Zirconium Deuteride Target

V.M. Bystritsky[1], A.P. Kobzev[1], A.R. Krylov[1], S.S. Parzhitskii[1], A.V. Philippov[1], G.N. Dudkin[2],

B.A. Nechaev[2], V.N. Padalko[2], F.M. Pen'kov[3], Yu.Zh. Tuleushev[3], M. Filipowicz[4],

Vit.M. Bystritskii[5], S. Gazi[6], J. Huran[6]

[1]Joint Institute for Nuclear Research, Dubna, Moscow Region, Russia

[2]National Research Tomsk Polytechnical University, Tomsk, Russia

[3]Institute of Nuclear Physics, Almaty, Kazakhstan

[4]Faculty of Energy and Fuels, AGH, University of Science and Technology, Cracow, Poland

[5]Trialpha Energy, Inc., Foothill Ranch, CA, USA

[6]Institute of Electrical Engineering SAS, Bratislava, Slovakia

E-mail: dudkin@tpu.ru

E-mail: bystvm@jinr.ru

**Abstract**

The mechanism for the *d(p,γ)*³He reaction in the region of ultralow proton-deuteron collision energies (6.67 <*E* <12.67 keV) is investigated using a target of zirconium deuterides. The experiment was carried out in the proton beam from the high-current pulsed Hall accelerator. Dependences of the astrophysical *S*-factor and the effective *pd* - reaction cross section on the proton-deuteron collision energy are measured. The results are compared with the available literature data. The results of this work agree with the experimental results obtained the LUNA collaboration with the target of gaseous deuterium.

## *Introduction*

Interest in the nuclear reaction

$$p + d \rightarrow {}^{3}\text{He} + \gamma \qquad (1)$$

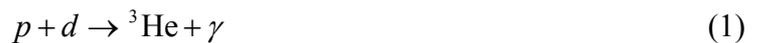

at ultralow proton-deuteron collision energies ($\sim$ keV) arises from the possibility of testing theoretical works on solution of three-body problems on the basis of modern concepts of nucleon-nucleon interaction potential within realistic two-body and two-body plus three-body forces [1-4], gaining information on the structure of exchange meson currents and determining their contribution to the *pd* interaction [5-7], checking and determining the amount of variation in the astrophysical *S*-factor for the *pd* reaction due to electron screening of the interacting particles [8], and solving some currently important problems in astrophysics [9-13].

From the standpoint of nuclear physics, it is particularly interesting to study the *pd* reaction in the region of ultralow proton-deuteron collision energies as a mirror reaction with respect to the reaction of radiative neutron capture by a deuteron [5, 7].

As to the role of the *pd* reaction in nuclear astrophysics, it should be mentioned that it is an essential reaction of the Big Bang nucleosynthesis as one of the main channels for production of $^4$He and it is also an important astrophysical reaction, the second step of the *pp* cycle governing the processes of evolution and nucleosynthesis in cold stars (in low-mass stars like our Sun, the *pp* cycle is one of the main (~ 98%) energy sources).

To know the *pd* reaction rate in the region of ultralow energies in the order of a few keV (the position of the Gamow peak for a protostar) is necessary both for calculating characteristics of evolution of stars and protostars (the *pd* reaction is the beginning of the protostar development process) and for verifying stellar evolution models. Deuterium, which is burnt is this case, is the primordial deuterium that resulted from the Big Bang rather than weakly generated in the *pp* interaction [12].

Until recently three experiments have been conducted [14-16] to investigate the dependence of the astrophysical *S*-factor for the *pd* reaction on the proton-deuteron collision energy (see Table 1).

Table 1. Experimental and calculated astrophysical $S_{pd}$-factors and linear approximation parameters for $S_{pd}(E) = S_0 + S_0' \cdot E$ ($E$ is the center-of-mass proton-deuteron collision energy).

| Ref. | $S_0$, eV·b | $S_0'$, eV·b·keV$^{-1}$ | $S_{pd}(0)$, eV·b |
|---|---|---|---|
| Viviani et al. [17] | | | $0.185 \pm 0.005$ |
| Griffiths et al. [14] | | | $0.25 \pm 0.04$ |
| Schmid et al. [15] | $0.166 \pm 0.005$ | $0.0071 \pm 0.0004$ | $0.166 \pm 0.014$ |
| Casella et al. [16] | $0.216 \pm 0.006$ | $0.0059 \pm 0.0004$ | $0.216 \pm 0.010$ |
| Bystritsky et al. [18] | | | |
| Quantity | $E$, keV | | |
| | 8.28 | 9.49 | 10.10 |
| $S_{pd}(E)$, eV·b | $0.237 \pm 0.0.61$ | $0.277 \pm 0.064$ | $0.298 \pm 0.065$ |

The *S*-factor for the *pd* reaction depends on the proton-deuteron collision energy in a practically linear manner, as should be expected from the analysis of the *s*-wave and *p*-wave contributions to the cross section for the radiative proton capture by a deuteron [14-15].

In [14-15] the investigated energy regions do not overlap the region of energies corresponding to the Gamow peak (3 ÷ 10 keV) for the *pd* reaction in a protostar and in the Sun, and extrapolation of the *S*-factor values to the zero proton-deuteron collision energy yields the result that differ by a factor of ~ 1.5.

The measured value of the *S*-factor for the *pd* reaction in the region of the Gamow peak [16] agrees with the results of extrapolating the experimental data [15] from the region of higher energies and is ~ 40% smaller than the result of extrapolating the experimental data [14] to the proton-deuteron collision energy region under study. The value of the *S*-factor for the *pd* reaction at the zero *pd* collision energy [16] agrees at the level of $3\sigma$ ($\sigma$ is the statistical error of the measurement of *S*) both with the calculated $S_{pd}$ [17] and with the results of extrapolating the experimental data [15] from the region of higher energies to the zero energy. The values of the *S*-factor obtained at the zero proton-deuteron collision energy by extrapolating the data from [15] and [16] differ by ~ 20%.

As is evident from the aforesaid, the previous reported results of investigating the pd reaction in the region of ultralow energies on classical accelerators were ambiguous and accordingly called for further investigation of this process using radically different methods for producing higher-intensity fluxes of accelerated hydrogen ions.

A method worth noting in this respect is the method of conducting the *pd* experiment using the high-current pulsed Hall accelerator with the closed electron current designed by us [19-22]. This accelerator produces beams of accelerated $H^+$, $D^+$, $^3He^+$, and $^4He^+$ plasma ions n the energy range of 2 to 20 keV [18-22]. The first experiments on measurement of the astrophysical *S*-factor and the *dd* reaction cross sections in the deuteron collision energy region of 2.2 to 6.3 keV [19-22] and the *pd* reaction cross sections in the proton-deuteron collision energy region of 8 to 10 keV [18] were carried out using the Hall accelerator. The results [18-22] indicated that it was a promising method for precise measurement of the characteristics of nuclear reactions in the ultralow energy region. The results [18] (see Table 1) ensured in a way the possibility of performing a more labor-consuming experiment on the study of the *pd* reaction (the yield of $\gamma$ rays from the *pd* reaction in the entrance-channel particle energy range of interest is on average four to 5 orders of magnitude lower than the yield of neutrons from the *dd* reaction for the corresponding deuteron collision energy interval) in a wider proton-deuteron collision energy range.

As to the electron screening of the interacting proton and deuteron, we can point out the following. In the adiabatic approximation [8] the expected increase in the value of the *S*-factor for the *pd* reaction is ~ 6% at the proton-deuteron collision energy 2.5 keV (in the center-of-mass system) and will amount to ~ 20% at the collision energy 1 keV. Nevertheless, as shows the investigation of the *dd* reaction mechanism in deuterides of metals, the conclusions drawn in [8] require experimental verification. For the $d(d, n)^3He$ and $d(d, p)^3H$ reactions at energies in the Gamow peak region the calculations in the adiabatic limit indicate that the electron screening insignificantly affects the behavior of the *S*-factor in the region of ultralow deuteron collision energies (the electron screening potential for the *dd* reaction in gaseous deuterium is $U_e = 27$ eV [8]). The experiments with deuterium-saturated metal targets showed however that in this case the electron screening potential $U_e$ increases by almost an order of magnitude [23-25] ($U_e = 200 \div 300$ eV and even 600 eV), which leads to considerable increase in the *S*-factor with decreasing deuteron energy in the energy region corresponding to the Gamow peak position. It is not impossible that screening of interacting protons and deuterons in deuterium-saturated metal targets similarly manifests itself for the *pd* reaction as well.

In this connection, the aim of this work was to measure the *S*-factor for the *pd* reaction in zirconium deuteride at the proton-deuteron collision energies ranging from 7 to 13 keV. Investigations of the *pd* reaction in ZrD₂ will also allow us to extract information on possible enhancement of this reaction due to the electron screening effect.

Below we describe the developed experimental setup and the results of investigating the *pd* reaction using solid-state deuteride targets and the pulsed Hall accelerator.

## *Measurement technique*

Experimental determination of the astrophysical *S*-factor and the effective *pd* reaction cross section in the region of astrophysical energies is based on measuring the yield of 5.5 MeV gamma rays from reaction (1) and using the parameterized dependence of the cross section for the reaction in question on the proton-deuteron collision energy

$$\sigma(E) = \frac{S_{pd}(E)}{E} \exp\left(-\frac{\beta}{\sqrt{E}}\right) \qquad (2)$$

For the *pd* reaction,

$$\beta = 31.29\sqrt{\mu}, \qquad (3)$$

$$N_\gamma^{tot} = N_p \varepsilon_\gamma \int\limits_0^\infty f(E) dE \int\limits_E^\infty \sigma_{pd}(E') n(x) \left(-\frac{dE'}{dx}\right)^{-1} dE' \qquad (4)$$

where $\mu$ is the reduced mass of the interacting particles in the entrance channel of the reaction in a.m.u., $E$ is the center-of-mass proton-deuteron collision energy, $S_{pd}(E)$ is the astrophysical *S*-factor for the *pd* reaction, $N_\gamma^{tot}$ is the total number of recorded gamma rays, $\sigma_{pd}(E)$ is the *pd* reaction cross section, $dE/dx$ is the specific proton energy loss in the target, $n(x)$ is the target deuteron density at depth $x$, $f(E)$ is the energy distribution function for the incident protons with average energy $E_p$, $\varepsilon_\gamma$ is the detection efficiency for the gamma rays from the *pd* reaction, and $N_p$ is the number of protons arriving at the target.

Parameterization (2) of the *pd* reaction cross section assumes interaction of "bare" protons with deuterons.

Considering the energy spread of deuterons incident on the deuterium target and the Coulomb proton energy loss resulting from interaction of protons with atoms (target molecules), the experimental values of the *S*-factor for the *pd* reaction are defined as [18, 26]

$$\overline{S_{pd}(E)} = \int_0^\infty S_{pd}(E)P(E)dE = S_{pd}(E_{col}) = \left. N_\gamma^{\exp} \middle/ N_p \varepsilon_\gamma \int_0^\infty f(E)dE \int_0^\infty \frac{e^{-2\pi\eta} n(x)}{E'(E,x)} dx \right. \tag{5}$$

where $2\pi\eta(E) = \beta / \sqrt{E}$,

$$P(E) = \frac{e^{-2\pi\eta} D(E) \int_E^\infty n(x(E,E')) f(E') dE'}{\int_0^\infty e^{-2\pi\eta} D(E) dE \int_E^\infty n(x(E,E')) f(E') dE'} \tag{6}$$

$$D(E) = -\frac{1}{E} \frac{dx}{dE} \tag{7}$$

$$E_{col} = \int_0^\infty E P(E) dE \tag{8}$$

$$\overline{S_{pd}(E)} \approx S_{pd}(\overline{E}) = S_{pd}(E_{col}) \tag{9}$$

where $N_\gamma^{\exp}$ is the number of recorded gamma rays from the *pd* reaction, $n(x)$ is the target deuteron density at depth $x'$, $E'(E_{cm}, x)$ is the proton-deuteron collision energy at depth $x'$, $P(E)$ is the distribution function for the probability of the proton-deuteron collision with subsequent recording of the gamma ray yield from the *pd* reaction, and $E_{col}$ is the proton-deuteron collision energy averaged over the distribution function $P(E)$.

At deep subbarrier energies, when the function $P(E)$ is a narrow peak, yield (4) can be calculated using the Laplace method and the expression for the gamma ray yield form the given reaction $N_\gamma$ can be written in a simple form in terms of the effective reaction cross section $\sigma_{pd}(E_{col})$ [26]

$$N_\gamma^{\exp} = N_p n_t \varepsilon_\gamma \tilde{\sigma}_{pd}(E_m) l_{eff}(E_m) K(E_m) \tag{10}$$

$$l_{eff} = \sqrt{\frac{2\pi}{-\varphi^{(2)}(E_m)}} \frac{dx}{dE}(E_m) \tag{11}$$

$$K(E_m) = \int_{E_m}^\infty f(E)dE = \frac{2E_m^{3/2}}{\beta} f(E_m) \tag{12}$$

where $l_{eff}(E_{col})$ is the effective range of the proton in the target, $E_m$ is the proton-target deuteron collision energy at the maximum of the function $P(E)$ $\varphi^{(2)}(E_m)$ is the second derivative of $\ln P(E)$ at $E_m$. The quantity $K(E_{col})$ defines the fraction of the initial proton flux contributing to the yield of gamma rays from the $pd$ reaction. At high energies $K(E_m) \to 1$.

As is evident from the above expressions, unambiguous information on the energy distribution and composition of the proton flux incident on the target, efficiency of the proton beam transport from the ion source to the target, probability for the proton beam neutralization during the transport, and deuterium concentration distribution over the target depth is needed for correct interpretation of the data obtained in the investigation.

## *Experimental setup*

The experiment on the measurement of the astrophysical *S*-factors for the *pd* reaction was carried out at the pulsed plasma Hall accelerator [18-22] at the average proton energies $E_p$ ranging from 11 to 19 keV. The energy distribution of the protons in the beam was described by the Gaussian function with an average FWHM spread of 16%.

Fig. 1 schematically shows the experimental setup. The pulse intensity of the accelerated proton beam was $5 \cdot 10^{14}$, the pulse duration was 10 µs, and the repetition rate was $\sim 5 \cdot 10^{-2}$ Hz.

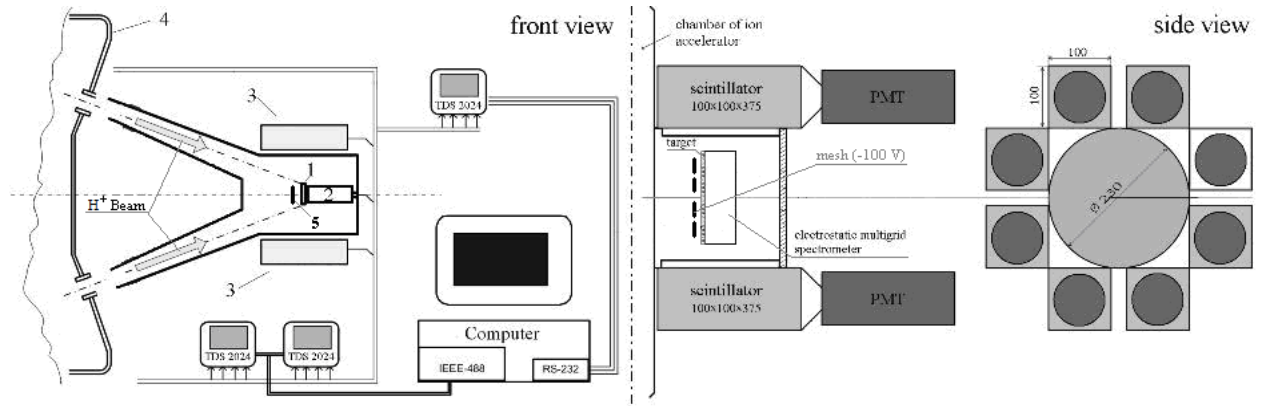

Fig. 1. Experimental layout: (1) solid-state $ZrD_2$ target, (2) multigrid energy analyzer, (3) gamma detectors base on NaI(Tl) crystals, (4) plasma Hall accelerator body, (5) grid.

Zirconium deuteride was used for targets. It was deposited on the stainless-steel backing by magnetron sputtering. The thus deposited zirconium deuteride layer was $\sim 1.5$ to 2 µm thick. The target diameter was 97 mm. Gamma rays with the energy $E_\gamma = 5.5$ MeV from the *pd* reaction were recorded by eight NaI(Tl)-based detectors (100×100×400 mm) placed around the target. The energy distribution of the incident protons was measured by a multigrid electrostatic spectrometer of charged particles, and the spatial distribution of the protons in the beam was measured by a linear set of Faraday cups placed along the radius of the target. In addition, for correct interpretation of the experimental data, a few parameters of the incident ion flux, such as the efficiency of the accelerated particle flux transport over a path of 300 mm from the ion source to the target in the range of angles from 0 to 20° and the beam composition, were measured by the time-of-flight method in special experiments.

The results of the experiments indicated that molecular hydrogen ion impurity of the accelerated proton beam was negligibly low ($\leq 1\%$) and the fraction of neutrals that resulted from charge exchange of the hydrogen ions on the residual gas in the measuring chamber of the accelerator during their transport from the ions source to the target was $\sim 1 \div 2 \%$ depending on the experimental conditions (initial composition of the ion beam from the Hall source and the partial pressure of the residual gas in the accelerator's measurement chamber which housed the $ZrD_2$ target).

The number of protons that interacted with the target in each accelerator pulse was found by integrating the target current. To suppress the electron emission from the target, a grid with a potential of -100 V (transparence 93%) was placed at a distance of 1 cm in front of the target.

The deuteron distribution over the target depth was measured by the elastic recoil deuteron (ERD) technique using the 2.3 MeV alpha-particle beam produced by the Van de Graaff accelerator [27-29]. In addition, together with detection of recoil deuterons, alpha particles scattered form the target nuclei to the backward hemisphere (RBS spectrum) were detected. The joint analysis of the ERD and RBS spectra allows the distribution of deuterons and impurity atoms over the target depth to be determined with a high accuracy. Also, using the Auger spectrometer [30] and a quartz balance, we determined the composition and thickness of the "parasitic" film on the zirconium deuteride target formed due to the effect of residual gas in the accelerator measuring chamber. The accelerator measuring chamber was evacuated using a zeolite adsorption roughing pump, a turbomolecular pump, and a cryogenic pump. The working vacuum in the measuring chamber was $\sim 10^{-7}$ mm·Hg.

To keep the "parasitic" layer on the target surface constant, dynamic equilibrium of the sorption and desorption from the target surface should be ensured, which requires continuous on-line monitoring of the condition of the adhesive layer on the target surface.

One of the ways to ensure on-line monitoring of sorption/desorption from the target surface is to measure the mass thickness of the precipitated/sputtered film using a quartz generator whose resonance frequency varies with the mass of the substance precipitated on the quartz plate (sensitivity up to $10^{-2}$ μg/Hz).

In our experiments we used a commercial Mikron-7 thickness meter with a frequency resolution of $\sim 1$ Hz. This instrument allowed the required sorption/desorption rate level to be maintained. On the whole, the results of the special test experiments indicated that the technique used for measuring the target surface purity allowed the system for evacuation of the accelerator measuring chamber and the volume of the accelerating ion diode to be monitored in real time for revealing and eliminating the factors contributing to contamination of the target surfaces due to the presence of residual gases.

These measurements are necessary for further analysis of the experimental data, where their result are use to consider correctly the proton energy loss to ionization in the "parasitic" layer until the protons reach the $ZrD_2$ layers.

The gamma detection efficiency of the experimental setup was determined by the Monte Carlo method. With the detecting equipment threshold of 3 MeV, it was $\varepsilon_\gamma = 0.300 \pm 0.006$.

This choice of the energy threshold was dictated by the necessity of suppressing the background of neutrons from the *dd* reaction that can be caused by the deuteron impurity of the proton beam (deuteron impurity of hydrogen was ~ $10^{-4}$).

The background from cosmic radiation and natural radioactivity was continuously measured during the experiment. To this end, background events were recorded between the working beam pulses for the time of 10 μs, which is equal to the duration of these working pulses of accelerated protons incident on the $ZrD_2$ target. The thus found background level was 12 to 1.5% for the proton energy range $E_p = 11 \div 19$ keV.

Fig. 2 presents the gamma energy spectrum obtained in one of the exposure with the zirconium deuteride target.

Spectrometric channels of the eight gamma detectors were calibrated using the standard $^{60}$Co, $^{137}$Cs, and Am-Be gamma-ray sources. The energy resolution measured with the $^{60}$Co source and averaged over all eight detectors was 4.3% for the 2.5 MeV total absorption peak line.

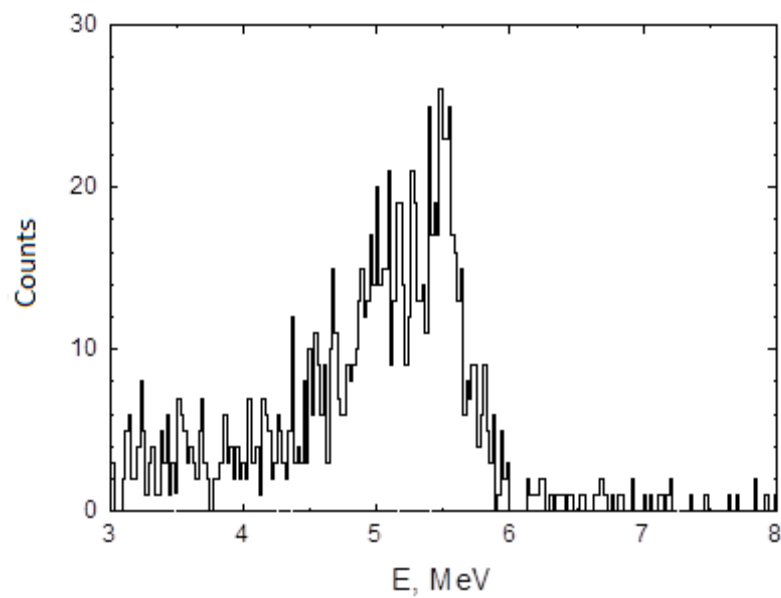

Fig. 2. Overall energy distribution of events recorded by eight NaI(Tl) detectors in the exposure with the $ZrD_2$ target at the proton energy 19 keV.

# *Experimental procedure and analysis of the experimental data*

The experiment included several exposures of the zirconium deuteride target to the accelerated proton beam. The *S*-factor for the *pd* reaction occurring in zirconium deuteride at the proton energies $E_p$ = 11÷19 keV was measured.

Table 2 lists the main data from the *pd* experiment. The astrophysical *S*-factor was determined by formula (5).

Table 2. Experimental data.

| Target | $E_p$, keV | $E_{col}$, keV | $N_p$, $10^{16}$ | $E_m$, keV | $S_{pd}$, eV·b (this work) | $S_0$, eV·b [16] | $S_0'$, eV·b·keV$^{-1}$ [16] | $\sigma_{pd}(E_m)$, $10^{-9}$ b |
|---|---|---|---|---|---|---|---|---|
| ZrD$_2$ | 11 | 6.306 | 192.2 | 6.808 | 0.246 ± 0.020 | | | 1.85 ± 0.21 |
| | 13 | 7.351 | 68.68 | 8.005 | 0.255 ± 0.020 | | | 4.07 ± 0.41 |
| | 15 | 8.383 | 77.2 | 9.196 | 0.265 ± 0.018 | 0.192 ± 0.048 | 0.0087 ± 0.0055 | 6.27 ± 0.44 |
| | 17 | 9.403 | 51.1 | 10.383 | 0.275 ± 0.018 | | | 9.08 ± 0.53 |
| | 19 | 10.413 | 32.3 | 11.570 | 0.281 ± 0.016 | | | 13.1 ± 0.93 |

Fig. 3 shows the dependence of the astrophysical *S*-factor for the *pd* reaction on the proton-deuteron collision energy in the interval 6.3 to 10.4 keV measured with the zirconium deuteride target in comparison with the similar dependences measured in [15, 16].

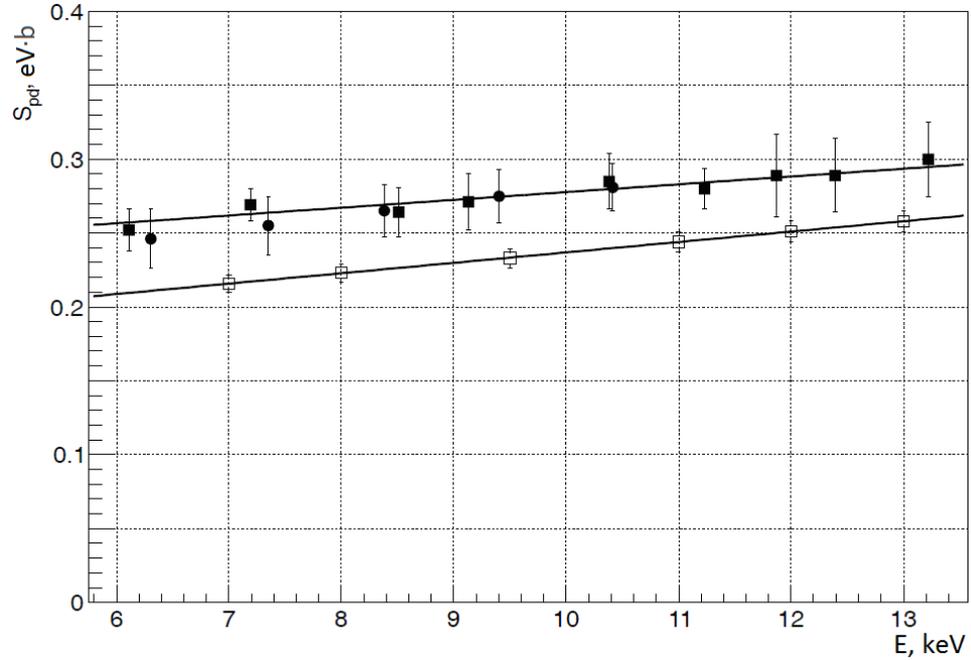

Fig. 3. Dependences of the astrophysical *S*-factor for the *pd* reaction on the proton-deuteron collision energy measured in this work (1), [16] (experiment with the gaseous deuterium target) (2), and [15] (experiment with the frozen heavy water target) (3).

As is evident from Fig. 3, our values of the astrophysical S-factor are in good agreement with the results from [16] obtained with gaseous deuterium and are greater than the results from [15] obtained with a target of heavy water ($D_2O$) for reasons that are still obscure.

The increasing linear dependence of the astrophysical *S*-factor on the proton-deuteron collision energy measured in [15-16] is confirmed within the measurement errors by the measurement in this work. Note that the statistical errors of measured $S_{pd}(E)$ and the limited energy interval of its measurements do not allow parameters of the linear functional dependence $S_{pd}(E) = S_0 + S_0'\cdot E$ to be determined with a high accuracy.

However, the results of this first experiment on measurement of the astrophysical *S*-factor for the *pd* reaction in the zirconium deuteride target do not differ within the measurement errors from the results of measurement in gaseous deuterium [16]. This indicates that the *pd* reaction enhancement due to the electron screening effect, if it ever exists, does not manifest itself in zirconium deuteride at the observable level. It thus follows that its influence on the intensity of the *pd* reaction in zirconium deuteride is much weaker than for the *dd* reaction. The theoretical evaluations support this conclusion.

To determine with a higher accuracy the absolute values of the *S*-factor for the *pd* reaction in deuterium-saturated metals and also the parameters $S_0$ and *pd* experiments should be conducted in a wider proton-deuteron collision energy interval at higher proton beam intensities.

Fig. 4 shows the dependence $\sigma_{pd}(E_m)$ of the effective *pd* reaction cross section on the proton-deuteron collision energy $E_m$.

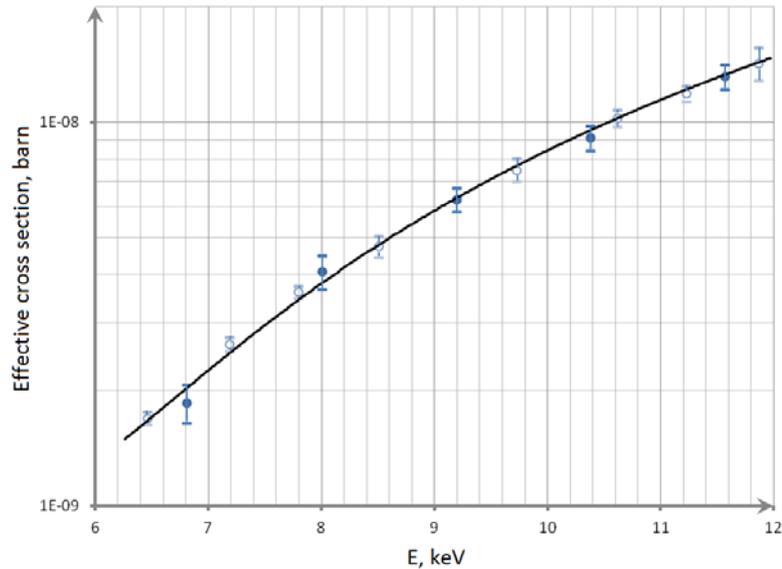

Fig. 4. Dependence of the effective *pd* reaction cross section on the proton-deuteron collision energy. Black circles are the results of this work obtained using formula (10), the solid curve is the calculation of the *pd* reaction cross section by formula (2) using the corresponding measured

*S*-factor values, and white circles are the effective *pd* reaction cross section values measured in [16].

It is seen that our dependence of the *pd* reaction cross section on the proton-deuteron collision energy calculated by formula (2) is in good agreement with the dependence of the experimentally measured effective *pd* reaction cross section on the proton-deuteron collision energy obtained using the measured gamma ray yields and equation (10).

The values of $\sigma_{pd}(E_m)$ were found from equation (10) using the experimentally measured yields of gammas from the *pd* reaction and the calculated values of the effective proton range in the target $l_{eff}(E_m)$ and $K(E_m)$. The thus obtained values of the effective *pd* reaction cross section in the proton-deuteron collision energy interval of 6.8 to 11.6 keV are presented in Table 2 and Fig. 4. The figure also shows the dependence of the *pd* reaction cross section on the proton-deuteron collision energy obtained using equation (2).

It follows from this result that the expression for the gamma-ray yield from the *pd* reaction can be represented in a simple analytical form by introducing the quantities $\sigma_{pd}(E_m)$, $l_{eff}(E_m)$ and $K(E_m)$.

In addition, Fig. 4 presents the experimental values of the effective *pd* reaction cross section measured with the gaseous deuterium target [16].

Good agreement between the results of this work and [16] in terms of extracting information on the value of the effective *pd* reaction cross section indicates that the analytical approach to determination of $\sigma_{pd}(E_m)$ considered by us in [26] is correct.

The authors are grateful to I.A. Chepurchenko for maintaining uninterruptable operation of the Van de Graaff accelerator and to E.I. Andreev for assistance in preparing the manuscript.

The work was supported by the Russian Foundation for Basic Research, project no. 12-02-00086-a, by the grant of the Plenipotentiary of Poland at JINR, in part by the Ministry of Education and Science of the Russian Federation, project no. 2.1704.2011 of the state task on research "Science", and in part by grant no. 2023/GF3 of the Ministry of Education and Science of the Republic of Kazakhstan.